# Flame spread over thin hollow cylindrical fuels in microgravity

Vipin Kumar[1,2], Amit Kumar[1*] Osamu Fujita[3], Yusuke Konno[3]

[1]Department of Aerospace Engineering, Indian Institute of Technology Madras, India.

[2] PhD scholar, Department of Aerospace Engineering, Indian Institute of Technology Madras

[3]Division of Mechanical and Space Engineering, Hokkaido University, Kita13Nishi8, Kita-ku Sapporo, Hokkaido, Japan.

*Corresponding author: Email: amitk@ae.iitm.ac.in, phone: +91-44-22574019

**Nomenclature**

| | |
|---|---|
| $C_s$ | : Specific heat of solid (1218 J/kg K) |
| $F$ | : Radiation view factor between two surfaces |
| $h$ | : Height of char column (mm) |
| $\Delta H_f^o$ | : Heat of combustion of cellulose fuel (15421 KJ/kg) |
| $K$ | : Constant for determining flame spread rate over planar fuel |
| $K_C$ | : Constant for determining flame spread rate over hollow cylindrical fuel (0.7) |
| $L$ | : length of hollow cylindrical fuel (cm) |
| $L_{ph}$ | : Preheat length (0.001065 m) |
| $\dot{m}_a$ | : Mass flow rate of air (kg/s) |
| $\dot{m}_f$ | : mass flow rate of fuel (kg/s) |
| $\dot{Q}_{cond}$ | : Conduction heat feedback (W/m$^2$) |
| $\dot{Q}_{rad}$ | : Radiation heat feedback (W/m$^2$) |
| $\dot{Q}_i$ | : Heat feedback on inner surface of fuel (W/m$^2$) |
| $\dot{Q}_{i,cond}$ | : Conduction heat feedback on inner fuel surface (W/m$^2$) |
| $\dot{Q}_{i,char\,rad}$ | : Radiation heat feedback from char to inner fuel surface (W/m$^2$) |
| $\dot{Q}_{i,surf\,rad}$ | : Surface radiation heat exchange at inner fuel surface (W/m$^2$) |
| $\dot{Q}_{i,amb\,rad}$ | : Radiation heat exchange from ambient to inner surface of fuel (W/m$^2$) |
| $\dot{Q}_{net}$ | : Net heat feedback on surface of fuel (W/m$^2$) |

| Symbol | Description |
|---|---|
| $\dot{Q}_o$ | : Heat feedback on outer surface of fuel (W/m²) |
| $\dot{Q}_{o,cond}$ | : Conduction heat feedback at outer surface of fuel (W/m²) |
| $\dot{Q}_{o,flame\,rad}$ | : Radiation heat feedback from flame to outer surface of fuel (W/m²) |
| $\dot{Q}_{o,surf\,rad}$ | : Surface radiation heat exchange at outer surface of fuel. (W/m²) |
| $\dot{Q}_{o,amb\,rad}$ | : Ambient radiation exchange from outer surface of fuel. (W/m²) |
| $R$ | : Radius of hollow cylindrical fuel (mm) |
| $T_{f,o}$ | : Flame temperature at outer region (K) |
| $T_C$ | : Temperature of char (800 K) |
| $T_{f,i}$ | : Flame temperature at inner core region (K) |
| $T_{so}$ | : Pyrolysis Temperature (670K) |
| $T_{s\infty}$ | : Fuel surface temperature at ambient. (298 K) |
| $T_\infty$ | : Ambient temperature (298 K) |
| $U_r$ | : Relative flow speed ($V_f + U_\infty$) |
| $U_\infty$ | : Opposed flow speed (cm/s) |
| $V_f$ | : Flame spread rate (mm/s) |
| $W$ | : Fuel width (mm) |

**Greek Symbols**

| Symbol | Description |
|---|---|
| $\alpha_g$ | : Diffusivity of gas, m²/s |
| $\alpha_s$ | : Thermal diffusivity of solid fuel (m²/s) |
| $\alpha_v$ | : Absorptivity of virgin fuel |
| $\delta$ | : Flame standoff distance (mm) |
| $\lambda_g$ | : Thermal conductivity of gas (0.04 W/mk) |
| $\epsilon$ | : Emissivity |
| $\epsilon_c$ | : Emissivity of char (0.9) |
| $\epsilon_v$ | : Emissivity of virgin fuel (0.8) |
| $\sigma$ | : Stefan Boltzmann constant (5.67E-8 W/m²K⁴) |
| $\rho_s$ | : Density of solid fuel (263 Kg/m³) |
| $\rho_g$ | : Density of gas (1.225 Kg/m³) |
| $\tau$ | ; Fuel thickness (0.0076 cm) |
| $\varphi$ | : Overall equivalence ratio |
| $\varphi_i$ | : Overall equivalence ratio at inner core region of fuel |
| $\varphi_o$ | : Overall equivalence ratio at outer region of fuel |

**Abstract**

This work presents experimental study on opposed flow flame spread over thin hollow cylindrical cellulosic fuel of diameters varying from 10 mm to 49 mm in microgravity environment. To understand the effect of flow and geometry on flame spread, experiments are conducted in low convective opposed flow conditions ranging from 10 cm/s to 30 cm/s for both hollow cylindrical and planar fuels at oxygen concentration of 21% and 1 atm pressure. In the microgravity environment the flame length and the flame spread rate are seen to increase with increase in hollow cylindrical fuel diameter over the flow range studied here. The flame spread rate exhibited a non-monotonic trend with flow speed, for flow of large diameter whereas a monotonic increasing trend is noted for small diameters. The flame spread rate over hollow cylindrical fuel is noted to be higher or at most equal compared to planar fuels over the matrix of experiments conducted in this study. A simplified analysis is carried out to arrive at an expression for flame spread rate over thin hollow cylindrical fuels. The analysis shows that the radiation heat transfer from the hot char to the inner surface of hollow virgin fuel dictates flame spread rate trend with fuel diameter of the hollow cylindrical fuels. Higher overall equivalence ratio in the inner section of the hollow fuels is responsible for higher char length in hollow fuels and also influence the flame spread rate for smaller fuel diameters.

**Novelty and Significance**

The present work is a first study on flame spread over thin hollow cylindrical geometry (circular ducts) in microgravity environments. Circular ducts are commonly encountered geometry and therefore the present study is important for fire safety in space. The geometry of hollow cylindrical fuel causes asymmetric of flames about the fuel and make flame spread mechanisms different from those for planar fuels and more strongly dependent on scale (fuel diameter). A simple analytical model is also developed to predict the flame spread rate with fuel diameter and elucidate the controlling mechanisms for the geometry.

**Authors Contributions**

Vipin Kumar: Experiments, data collection and writing.

Amit Kumar: Writing, editing and supervision.

Osamu Fujita: Review and technical discussion.

Yusuke Konno: Review and technical discussion.

**1. Introduction**

Flame spread over solid combustible material such as cellulose and other polymers has been studied over five decades out of concern for better fire safety. This research area has another dimension in the present day due to rapid increase in manned space missions and space travel [1], [2]. In the spacecrafts the gravity level is near zero. Further to maintain adequate air-conditioning and ventilation, there is always a small convective flow inside the spacecrafts which is of the order of a few cm/s [1]. The oxygen concentration and pressure inside cabin may also vary[1]. Therefore, it is important to understand flame spread mechanism over solid surface in these conditions. Flame spread over solid surface is a complex process which involves momentum, energy and species transport along with complexity of solid pyrolysis and gas phase reactions. Early attempts to perform experiments over thin solid fuel in reduced gravity were made by Kimzey et al.[3] and Neustein et al.[4] in aircrafts flying in Keplerian trajectories. However, due to high g-jitters no major conclusion on low gravity flame spread rates could be made comparing with those of normal gravity environment. A series of experiments using ground based drop tower facility were performed [5], [6], [7] to study the flame spread behavior over planar thin solid fuel in microgravity and a detailed flammability map over thin solid fuel is constructed with molar concentration of oxygen and characteristic flow velocity as co-ordinates. Fujita et al. [8][9] performed experiments on downward flame spreading over the electric wires in microgravity and reported that flame spread rate in microgravity is higher as compared to normal gravity environments in similar experimental conditions. The higher flame spread rate in microgravity compared to normal gravity is also reported for cylindrical PMMA rods at different oxygen concentration and opposed flow speed [10]. The data shows a non-monotonic trend in flame spread rate with respect to opposed flow speed at each oxygen concentration. Large scale flame spread experiments is also conducted in orbiting spacecraft environments to understand scale effect on flame spread in microgravity[11].

Physics based theoretical analysis of flame spread was reported for the first time by de Ris [12]. Flame spread over thin and semi-infinite solid fuel bed was theoretically analyzed and expressions for flame spread for each case was obtained by solving the energy and species conservation equations. The model assumed infinite chemical kinetics in the gas phase. The effect of finite kinetics on the flame spread rate was presented in work of Fernandez-Pello and Hirano[13] where an expression using thermal theory and Damköhler number was proposed. Bhattacharjee et. al. [14] obtained an expression for flame spread rate and flammability limit in microgravity environment using a simplified analysis. They also correlated the flame geometry in opposed flow environments[15]. Several investigators [16], [17], [18], [19] have numerically modeled the flame spread mechanisms by solving the gas phase and solid phase conservation equations along with solid fuel pyrolysis in opposed flow, normal gravity and microgravity environments.

Over the years flame spread experiments have mostly been on planar fuels or cylindrical rods and wires. Additionally, experiments and numerical simulations on multiple parallel fuels[20], [21], [22], [23], parallel plates[24], solid fuel tubes [25], [26] are investigated to understand flame spread behavior with adjacent flames nearby. A normal gravity experimental study [20], reported that flame spread rate increases with increase in number of fuel sheets and remains almost constant for more than seven fuel sheet. A numerical study in microgravity environment [21] and experiments in normal gravity [22] showed a non-monotonic trend in spread rate with spacing between fuel sheets and difference in flammability behavior compared to single fuel sheet was also reported. Joshi et al.[23] have simulated flame spread behavior over multiple parallel fuels with different spacing between fuels and found the non-monotonic trend in spread rate which is due to competition between oxygen transport and radiation heat transfer to fuel surface.

Even though flame spread behavior over thin cellulosic fuel has already been studied extensively, the flame spread behavior over hollow cylindrical fuel in microgravity has not been reported yet. In fact, no studies have been focused on effect of hollow cylindrical geometry on flame spread rate in microgravity environments. However, an attempt was made to describe flame spread mechanism over hollow cylindrical geometry in normal gravity [27] experimentally and predicted that flame spread rate increases with fuel diameter in normal gravity environments.

In this paper we investigate the effect of diameter on flame spread rate over thin hollow cylindrical cellulosic fuel in microgravity environment and how flame spread behavior over hollow cylindrical geometry is different from the flame spread over thin planar fuels. A theoretical analysis is also presented to explain the experimental results and arrive at simple model equation for describing flame spread behavior over hollow cylindrical geometry.

## 2. Experimental Apparatus and procedure

### 2.1. Experimental module

The microgravity experiments are performed using 2.5 s drop tower facility available at National Centre for Combustion Research and Development (NCCRD), IIT Madras, India. The schematic of drop tower is shown in Appendix Fig. A1. Further details on the design and operation of this drop tower facility can be found in the literature[28]. Here in brief an experimental module consists of rectangular duct of 12×12 cm inner cross section. A small DC fan is placed at the bottom of duct which provides the varying opposed flow speed. Honeycomb and fine mesh are placed in-between fuel holder and fan to obtain a uniform desired opposed flow in the test section. The cross-sectional view of duct is shown in Fig. 1 (a). The calibrated opposed flow speed values are shown in Appendix Fig. A2. The top section of duct is made of transparent acrylic to observe the combustion phenomena over the solid fuel. The fuel sample is held at centre of the rectangular duct with the help of different thin metallic fuel holders as shown in Fig. 1(c). A polycarbonate enclosure with capacity of 40 litre is used to enclose the duct assembly (Fig. 1(b)) so that the ambient conditions inside chamber can be varied. However, the present experiments are done at atmospheric oxygen concentration of 21% and pressure 1 atm. Two digital camera is mounted over outer wall of transparent enclosure. The whole experimental module is fixed over top deck of inner capsule, whereas bottom decks of inner capsule have provision for data acquisition models and power supply unit (Fig. 1(d)).

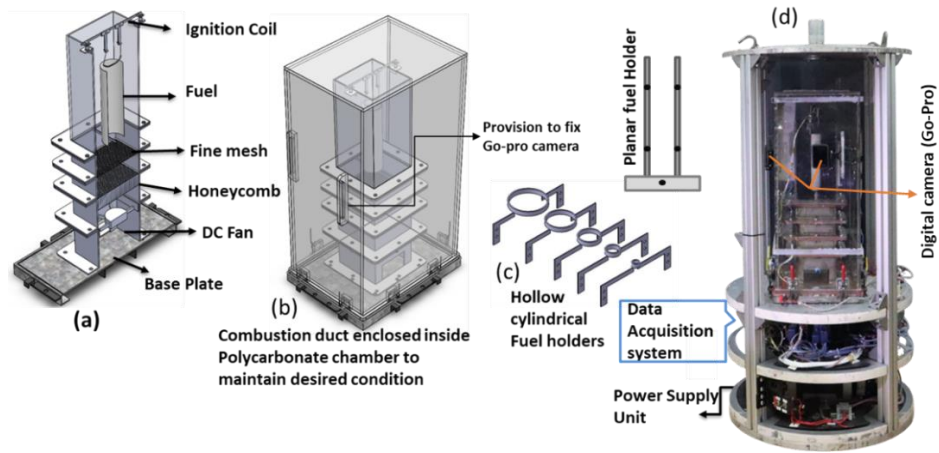

**Fig. 1.** The experimental module **(a)** Sectional view of combustion duct **(b)** Experimental set up enclosed with polycarbonate enclosure **(c)** Fuel sample holders **(d)** Complete experimental module placed over inner capsule.

## 2.2. Fuel Samples

Because of limited time (2.5 s) available for microgravity environment using drop tower facility, it is better to select a fuel over which flame spread can quickly achieve a steady state condition with rapid flame spread rate. Here the fuel used in the experiment is 76-μm thick cellulosic paper (Kim-wipes) with composition of 99 % cellulose and 1% polyamide. The area density of this paper is about 18 gsm (grams per square meter). The height of fuel sample is fixed to 180 mm which can be fitted inside the combustion duct, with different hollow cylinder diameters ranging from 10 mm to 49 mm. The maximum limit of diameter (49 mm) is selected to ensure the shape of the hollow cylinder and cross section remains same with ignition coil and during the flame spread process. For making hollow cylindrical fuel, paper is wrapped around a metallic tube of desired diameter and both edges are stuck at few points using small drop of glue at few points on the paper edges and then directly slide over hollow fuel holders. The hollow fuels stand over 1.5 mm thick stainless steel fuel holders (fig. 1(c)) and to ensure the samples remain straight, the bottom edge is stuck to the fuel holder using an adhesive tape. For planar fuel samples, the fuel is secured in between the arms flat fuel metallic sample holders

(fig. 1(c)) up to a length of 150 mm. Extra section of 30 mm length is free over the ignition coil for igniting the fuel. The widths of the fuel are selected as 10 mm, 20 mm and 40 mm.

## 2.3. Ignition system

The fuel is ignited with Kanthal wire coil having coil diameter of 4 mm and wire diameter of 0.4 mm. The wire is selected because it heats and cools quickly and ignite the fuel sample smoothly. About 20 mm to 30 mm of fuel sample is exposed to the ignition coil in the combustion duct. This coil (about 5 $\Omega$ resistance) draws 4 A of current at 24 V DC power for 0.3s to 0.5 s to ignite the cellulose fuel samples. It can be noted that in the present experiments the fuel samples are ignited in normal gravity and once fuel is ignited the ignition coil power is switched off and then dropped for 2.5 s to obtain flame spread data in microgravity environments.

## 2.4. Imaging and data processing method

The flame spread behaviour of different specimen is recorded with the help of two digital camera (Go-Pro Hero-10) fixed over the polycarbonate enclosure from two perpendicular sides (Fig. 1 (b)). The cameras are set at 120 FPS and 1.2 K resolution for better imaging of spreading flames. A white LED light is provided in the background to uniformly illuminating the fuel samples. The purpose of uniform illumination of fuel is to avoid the light reflections coming from the flame to the white surface of fuel and to identify the clear leading edge of spreading flames on fuel surface.

The data obtained from the camera is post processed using MATLAB code. Method of tracking the flame leading edge over the solid surface will help in calculating the flame spread rate. To make the distinction in intensity of flame and fuel surface, the uniformly illuminated, intensity equal or higher than that of the white part of the fuel is cropped from all the frames using MATLAB crop function. The cropped image is converted into gray image and threshold intensity corresponding to pyrolysis front over 100 pixel width of the fuel is selected. The position of this threshold intensity (corresponding to flame leading edge) is tracked along the length of fuel in each frame. The data of position of flame leading edge with time is obtained. A typical data corresponding to flame spread over 10 mm and 49 mm hollow cylindrical fuels is shown in Fig. 2. One can note that the variation of flame position with time is linear. The average flame spread rate is taken as the slope of the least squared error linear fit of position vs time plot.

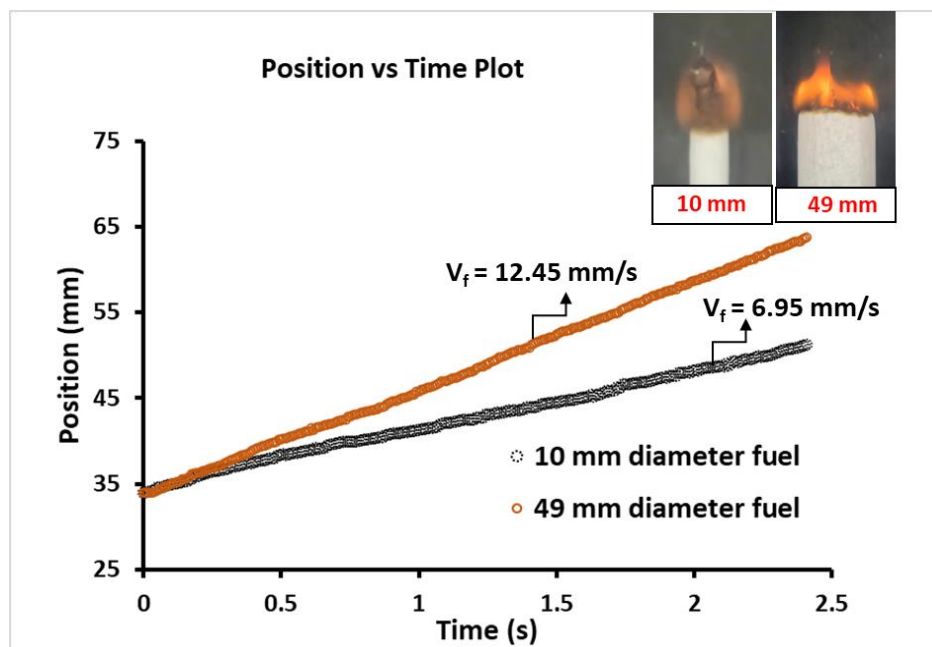

Fig. 2. Variation of position of flame leading edge with respect to time for 10 mm and 49 mm diameter fuel at 20 cm/s opposed flow speed.

It can be seen in plot that the fuel diameter 49 mm has higher slope as compared to 10 mm fuel diameter which correspond to flame spread rates of 12.45 mm/s and 6.95 mm/s respectively.

3. **Experimental Results**

Experiments are carried in microgravity environments at 21% $O_2$ and 1 atm. pressure for hollow cylinder fuel samples of diameters 10 mm, 19 mm, 29 mm, 38 mm and 49 mm. In addition to this, for comparison, experiments are carried out for planar fuel of widths 10 mm,

20 mm and 40 mm. The opposed flow speed was varied from 10 cm/s to 30 cm/s in steps of 5 cm/s. The lower flow speed limit is set to 10 cm/s due to starting voltage required to overcome inertia of the DC fan and upper limit was chosen close to the magnitude of induced buoyant speed over the thermal length in normal gravity. For each drop test in the above set of experiments, the flame spread process in micro-gravity is video recorded and flame spread rates are determined. Following this, the effect of fuel size and geometry on flame spread rate at various opposed flow speeds are examined and discussed.

### 3.1. Visible flame characteristics in microgravity

The flame is quick to respond to sudden change in gravity level at the start of the drop test. Figure 3 compares the instantaneous images of spreading flame over hollow cylindrical fuels of 29 mm diameter under various flow speeds. Note that fuel is exposed to external lighting to track pyrolysis front. In absence of external illumination flame in image of Fig.3 will be more prominently visible which is shown in Appendix A, Fig.A3. In Fig.3. the images in first column show downward spreading flame in normal gravity and remaining columns show snapshots of spreading flames in microgravity at the start of the drop (0+ s) and at time intervals of 0.35 s up to the end of the drop test. It may be noted from Fig. 3 that for all flow speeds the flame length (i.e., size of the flame along the spread direction) increases with the passage of time in micro-gravity. This is due to reduced convection in absence of buoyancy in microgravity, where the transport of fuel vapor and oxidizer are increasingly more affected by slow diffusion process and hence longer flame. Consequently, the flames are also less bright in micro-gravity [7]. The flame leading edge which is located on the side of the fuel surface in normal gravity is also seen to shift downstream with larger standoff distance from the virgin fuel in microgravity. The influence of external flow speed on flame appearance can also be noted in Fig. 3, over the range of flow speeds studied here. The brightness of flame progressively increases at higher values of opposed flow speeds. The increase of flow speed is known to increase oxygen supply to the flame and hence the flame temperature in micro-gravity [9] , and could be the cause for increase in flame intensity noted here at higher flow speeds.

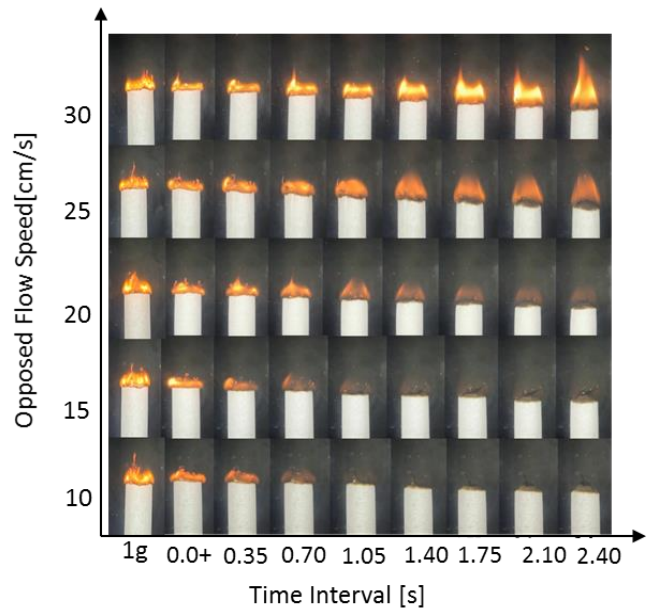

Fig. 3. Instantaneous images of spreading flame over 29 mm hollow cylindrical fuel at different opposed flow speeds (Vertical axis) and at normal gravity, start of drop test and time intervals of 0.35 s after the drop (Horizontal axis)

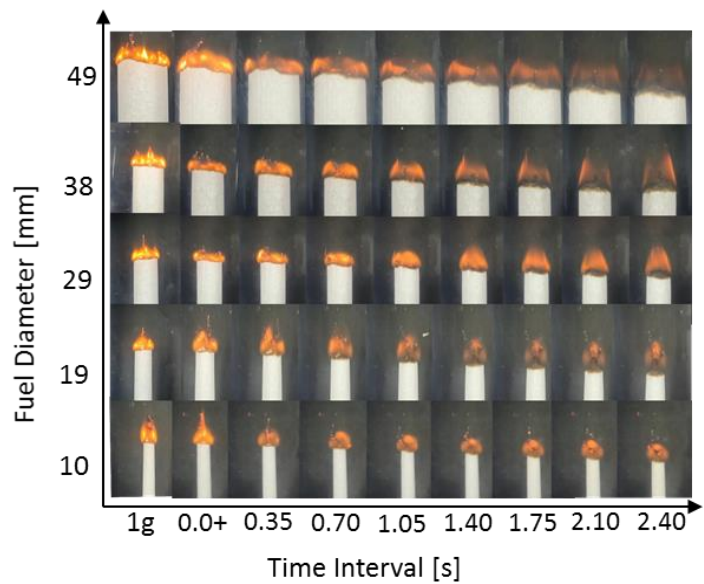

Fig. 4. Snapshot of spreading flame over different diameters of hollow cylindrical fuel (vertical axis) with respect to time interval of 0.35 s after drop (Horizontal axis) at 25 cm/s opposed flow speed.

The flame shape over hollow cylindrical fuels of various diameters and at opposed flow speed of 25 cm/s is shown in Fig. 4. Similar to Fig. 3, the first column shows a normal gravity flame and remaining microgravity flames from start to end of the drop test at interval of 0.35 s. As noted before in Fig. 3, in microgravity, as time progresses, the length of flame increases and luminosity decreases for all fuel diameters. It is interesting to note that the flame size varies with diameter for hollow cylindrical fuels. In normal gravity the flame is about 25 mm tall for fuel diameter of 10 mm and becomes 15 mm tall for fuel diameter of 38 mm. This could be because of increase in oxygen availability with the increase in fuel diameter in the possibly under ventilated in the inner regions of fuel. The micro-gravity flame in contrast, for 10 mm diameter fuel is little shorter at about 2 cm (but wider than normal gravity flame) and for fuel diameter 38 mm the flame is as long as 3.2 cm. It is interesting to note that while flames are generally longer (Fig. 3 and Fig. 4) in micro-gravity, for 10 mm fuel diameter, the flame length is shorter in microgravity. We can also note that the flame length in micro-gravity increases with increase in fuel diameter, a trend opposite of normal gravity. This is indicative of increased flame spread rate due to some effect which overshadows the effect of increased ventilation in larger fuel diameter. It is also interesting to note the variation of char height with fuel diameter. For the fuel diameter of 10 mm the char is long and exceed the flame height both in normal gravity and micro-gravity. As the diameter of the fuel increases the height of char column decreases. One can note change in flame shape with fuel diameter from the images in the right most column. For small fuel diameters (here 10 mm and 19 mm), the flame is seen surrounding the char layer, for fuel diameter of 29 mm, a candle like flame with tip at the axis of fuel is seen and for large fuel diameters (here 38 mm and 49 mm) an annular flame is noted. The different shapes of flame may be related to oxygen availability to burn the fuel vapor in the core region of the cylindrical fuels. As the core area increases more oxygen is available to burn fuel and oxidize the char and thus reduce the char length.

## 3.2. Flame spread rate variation with fuel diameter

The rate of flame spread over hollow cylindrical fuels is determined from the motion of the flame front along the fuel captured in video by two perpendicularly placed digital cameras (see Fig. 1(d)). As explained before, the video is decomposed into frames of instantaneous images of the spreading flame. The position of the flame front is tracked with time to determine the flame spread rate. Figure 5(a). shows the variation of flame spread rate with respect to fuel diameter at different flow speeds obtained in the microgravity experiments. The flame spread rate over hollow cylindrical fuels is seen to increase monotonically with the increase in fuel diameter. While this trend is true for all the five flow speeds in this study, the flame spread rate curves of some flow speeds cross each other. For example, for fuel diameter of 10 mm the highest flame spread is for $U_\infty = 30$ cm/s and least for $U_\infty = 10$ cm/s. However, for fuel diameter beyond 29 mm, the highest flame spread rate is for $U_\infty = 20$ cm/s and lowest for 10 cm/s. This implies that the flame spread rate has a non-monotonic trend with opposed flow speed. This is discussed in the next sub section.

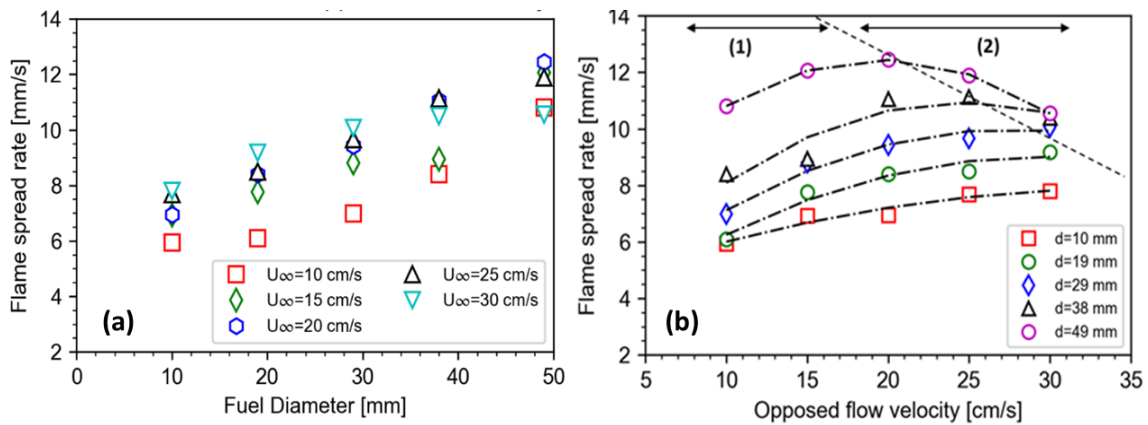

Fig. 5 (a). Variation of flame spread rate with fuel diameter and (b) Opposed flow speed of thin hollow cylindrical fuel in microgravity environment. (Dashed line in (b) shows point of change from regime (1) to regime (2))

### 3.3. Effect of opposed flow speed on flame spread rate

The data in Fig. 5(a) is replotted in Fig. 5(b). to show variation of flame spread rate with opposed flow speed for a fixed hollow cylindrical fuel diameter. It is noted that over the flow speed range experimented in this study, the flame spread rate over 10 mm, 19 mm and 29 mm fuel diameters monotonically increases. However, the flame spread rate for 38 mm and 49 mm fuel diameter shows a non-monotonic trend with respect to opposed flow speed with a maximum at certain flow speed for a given diameter. The non-monotonic increase-decrease trend of flame spread rate has been widely reported in the literature for planar fuels [6], [7], [29] as well as cylindrical fuels [9]. The increasing-decreasing trend is usually split in two regimes about the critical flow speed at which flame spread is maximum. For flow speeds less than this critical flow speed, here referred to as Regime 1, where the flame spread rate increases with opposed flow speed, has been referred in literature as oxygen transport regime [7] or radiation regime [30]. In regime 2, the flame spread rate decreases as the opposed flow speed increases. This regime has been accepted by researchers to be due to shorter flow residence time compared to the reaction time (small Damköhler number). In the present experiments, the critical flow speed at which flame spread reaches a maximum is seen to shift to lower flow speeds as the fuel diameter is increased. This is clearly seen in case of 38 mm and 49 mm fuel diameters.

## 3.4. Flame spread rate over hollow cylindrical fuels compared to planar fuels

To understand better the phenomena of flame spread over hollow cylindrical fuels a few experiments are also carried out for planar fuels. Experiments are carried out for planar fuel of widths 10 mm, 20 mm and 40 mm for five flow speeds in the range on 10 cm/s to 30 cm/s. The flame spread over two different shapes viz planar and hollow cylindrical fuels are compared in microgravity. Figure 6 shows comparison of flame shapes over hollow cylindrical fuel of diameters 10 mm, 19 mm and 38 mm in normal gravity and microgravity along with flame shapes over corresponding planar fuels of width 10 mm, 20 mm and 40 mm. The flow speed is 25 cm/s and oxygen level is 21%. The micro-gravity images are at instant close to the end of the drop experiment (~ 2.47 s). For planar fuels, both the front view and the side view is shown.

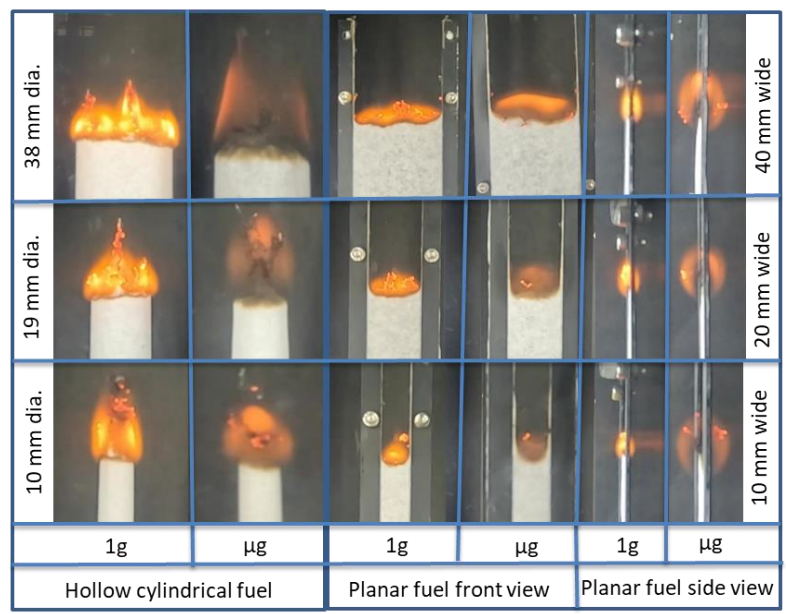

Fig. 6. Comparison of flame shape for flame spreading over cylindrical and planar fuels in normal gravity and microgravity at opposed flow speed of 25 cm/s.

It may be noted that unlike for cylindrical fuels, where the flame size varies with diameter, flame size is almost not affected by fuel width for planar fuels. The flame length and thickness for the planar fuel are about 1 cm and 0.5 cm respectively in normal gravity which increases to about 2 cm and 1 cm respectively in microgravity for all three fuel widths. However, for cylindrical fuels as mentioned before, the flame length in normal gravity decreases with diameter whereas flame length increases with diameter in micro-gravity. The char is more prominent in hollow cylindrical fuels compared to planar fuels. The char length is seen to increase in micro-gravity for planar fuels but is seldom greater than flame height. The common feature of spreading flame over planar and cylindrical fuels in micro-gravity is that in both cases flame is less luminous compared to the normal gravity flame.

The flame spread rates were determined for various tests on planar fuels following the procedure described for the cylindrical fuel. Figure 7 shows a comparison of the flame spread rate for the hollow cylindrical fuel and the planar fuels of corresponding size for various flow speeds. The flame spread rates for cylindrical fuel diameter of 10 mm and those of planar fuel of width 10 mm are nearly the same for the entire flow speed range (Fig. 7(a)). In Fig. 7 (b) flame spread rate over 19 mm diameter fuel with 20 mm wide planar fuel are compared. The flame spread rate is almost similar at smaller opposed flow speeds, but once opposed flow speed increases, the flame spread rate for cylindrical fuel becomes much more compared with planar fuel. Finally for large diameter of fuel, here in case of 38 mm diameter the flame rate always higher as compared with planar fuel flame spread rates (Fig 7(c)). In contrast to hollow cylindrical fuel, for the planar fuel, as the width of the fuel is increased from 20 mm to 40 mm, there is no significant change in the flame spread rate. Itoh et al.[27] showed that radiation transport may greatly influence the flame spread rate in hollow cylindrical fuels in normal gravity and also they concluded that above 10 mm diameter of fuel flame spread rate is always greater than planar fuel in normal gravity.

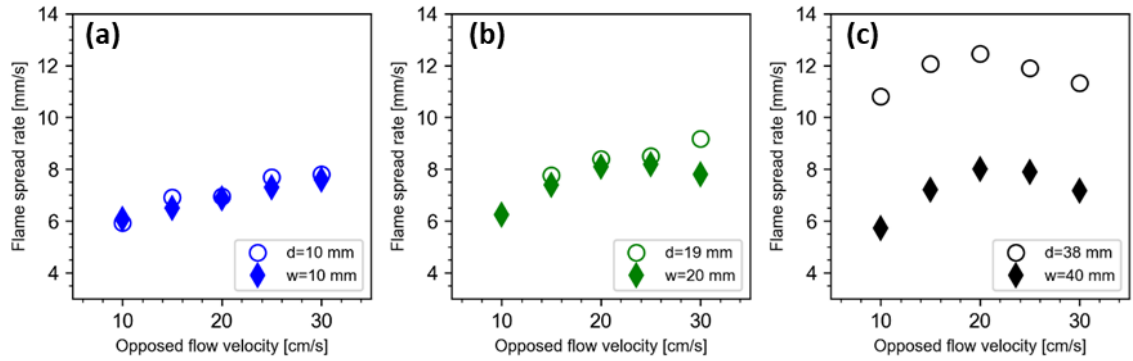

Fig. 7. A comparison of flame spread rate over planar having 10 mm, 20mm and 40 mm wide fuel and hollow cylindrical fuel having 10 mm ,19 mm and 38 mm fuel diameter in (a), (b) and (c) respectively.

## 4. Analysis of flame spread rate over hollow cylindrical fuel

In the above section it was seen that flame spread process over hollow cylindrical fuel in micro-gravity in many ways different from the flame spread process over the planar fuels. The flame shape and flame spread rates show a strong dependence on hollow cylindrical fuel diameter, whereas the planar fuel have very weak dependence on fuel width. Formation of char is also seen increased in hollow cylindrical fuels compared to the planar fuels. In order to understand the factors affecting flame spread process, a simplified analysis is carried out here for both planar and hollow cylindrical fuels.

The earliest physics-based analytical model on opposed flow flame spread was proposed by de Ris [12] for both thermally thin and thermally thick semi-infinite fuels. The model assumes infinite gas phase kinetics and gives an expression for flame spread driven primarily by heat feedback to the virgin solid fuel from the flame by conduction and radiation. This is often referred to as the thermal model. The flame spread rate model has no dependance on the opposed flow speed. Over the years the model has been revisited by many researchers [31], [32], [33], for improved flame spread prediction. In the present study the fuel is a thermally thin fuel, where thermal conduction time scale across the fuel, $t_{cond,s} \sim \frac{\tau^2}{4\alpha_s} (\approx 0.004) \ll t_{spread} \sim \alpha_g/U_r V_f (\approx 0.13)$, the flame spread time scale. In such a situation heating of solid fuel is primarily from the gas phase and temperature is uniform within the depth of fuel. Figure 8 (a) shows a schematic of opposed flow flame spread where a laminar flame heats the virgin fuel ahead of the flame leading edge by conduction and radiation. The steady laminar flame spread rate over fuel of finite width, 'W' can be obtained from the energy balance across a control volume (Fig. 8(b)) ahead of the flame leading edge in the preheat region. In this region there is almost no pyrolysis of fuel, solid fuel is only preheated from ambient temperature, $T_{s\infty}$ to high pyrolysis temperature, $T_{so}$ due to heat conduction, $\dot{Q}_{cond}$ and net radiation, $\dot{Q}_{rad}$ at the fuel surface. Net radiation ($\dot{Q}_{rad}$) consists of radiation feedback from the flame to solid surface and radiation loss from the solid surface to ambient.

$$\dot{Q}_{cond} + \dot{Q}_{rad} \sim \rho_s V_f \tau W C_s (T_{so} - T_{s\infty}) \tag{1}$$

$$\text{or} \quad V_f \sim \frac{(\dot{Q}_{cond} - \dot{Q}_{rad})}{\rho_s \tau C_s W (T_{so} - T_{s\infty})} \tag{2}$$

here $\dot{Q}_{cond} \sim \lambda_g \frac{(T_f - T_{so})}{L_{ph}} L_{ph} W$ and $\dot{Q}_{rad} \sim \epsilon \sigma (T_{so}^4 - T_\infty^4) L_{ph} W$ are heat transfer by conduction and heat transfer by radiation respectively. In these expressions, $\lambda_g$ is thermal conductivity of gas, $T_f$ is flame temperature, $L_{ph}$ is the preheat length scale and the flame standoff distance, $\rho_s$ is the density of solid, $V_f$ is flame spread rate, $\tau$ is the fuel thickness, flame width 'W' is same as the fuel width, $C_s$ is specific heat capacity of solid fuel and $T_{SO}$ and $T_{s\infty}$ are the pyrolysis temperature and unburnt solid fuel temperature respectively. Therefore, Eq. (2) can now be written as

$$V_f = K \frac{\lambda_g (T_f - T_{so}) - \epsilon \sigma (T_{so}^4 - T_\infty^4) L_{ph}}{\rho_s \tau C_s (T_{so} - T_{s\infty})} \tag{3}$$

Here, K is a constant. The value of constant has varied in the analysis by different researchers [12], [32], [34]. Here we adopt a similar procedure to obtain an expression of opposed flow flame spread rate over hollow cylindrical fuel in micro-gravity. A schematic of opposed flow flame spread over a hollow cylindrical fuel is shown in Fig. 8(c). Unlike planar fuel where flames on either side of the fuel are nearly identical, in case of cylindrical hollow fuel the flame extensions on two sides of the fuel are asymmetric. This is because of two differently oriented curved surfaces of the hollow cylinder. On the inner surface the flame on a small segment of fuel interacts with flames over neighbouring segments whereas flame on the outer surface is exposed only to the ambient. Further, depending on the opposed flow speed and flame spread rate the availability of oxidizer in the inner region of the hollow cylinder may be limited or insufficient to react all fuel pyrolyzed from the fuel surface. The outer surface on the other hand is always over ventilated due to large flow cross section between the fuel and the test tunnel.

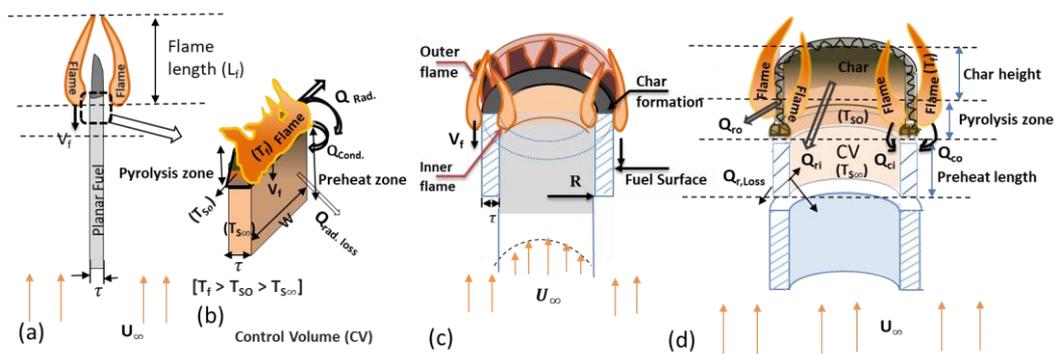

Fig. 8. Conceptual schematic of flame spreading over planar and hollow cylindrical fuels, (a) planar fuel (b) control volume over preheat region for planar fuel (c) sectional view of flame spread along length of hollow cylindrical fuel (d) Energy transport processes in control volume over preheat region (CV) of hollow cylindrical fuel.

In a procedure similar to the planar fuel, considering heat balance over a control volume in the preheat region of a hollow cylindrical fuel (Fig. 8 (d)), the opposed flow flame spread rate, $V_f$ along the length of the fuel can be expressed as

$$V_f = K_c \frac{\dot{Q}_o + \dot{Q}_i}{\rho_s \tau C_s . 2\pi R(T_{so} - T_{s\infty})} \quad (5)$$

In equation (5), $K_c$ is empirical constant for obtaining actual flame spread (here $K_c = 0.7$ is considered) and value is similar as in case of planar fuel. The heat feedback rates are $\dot{Q}_i$ and $\dot{Q}_o$ to the preheat region (the control volume) on the inner and the outer surfaces of the fuel respectively, are different unlike in the case of planar fuel. As before, $\rho_s$ is virgin fuel density, $\tau$ is the fuel thickness, $C_s$ is the specific heat capacity of the virgin fuel, R is the radius of the fuel and $T_{so}$ and $T_{s\infty}$ are the virgin fuel temperatures namely, pyrolysis temperature and initial temperature respectively.

Heat feedback $\dot{Q}_o$ and $\dot{Q}_i$ include contribution due to both conduction and radiation. $\dot{Q}_o$ comprises of conduction to preheat region, $\dot{Q}_{0,cond}$ from the flame leading edge, radiation contribution flame, $\dot{Q}_{o,flame\ rad}$ and net radiation loss from fuel surface, ($\dot{Q}_{o,surf\ rad} - \dot{Q}_{0,amb\ rad}$). Since the view factor of char to the outer surface of the fuel is zero, there is no radiation contribution to the preheat region of the fuel outer surface. Further, since flames are small and thin, flame radiation, $\dot{Q}_{o,flame\ rad}$ from the flame is also negligibly small [23], [27] $\dot{Q}_{0,flame\ rad} \rightarrow 0$. Therefore,

$$\dot{Q}_o = \dot{Q}_{o,cond} - (\dot{Q}_{o,surf\ rad} - \dot{Q}_{o,amb\ rad}) \quad (6)$$

The components of $\dot{Q}_i$ at the inner surface of fuel is somewhat more complex. This is because concave inner surface is a self-facing surface where radiation from hot char, $\dot{Q}_{i,char\ rad}$ also contributes to radiant heating of the fuel in the preheat region (Fig. 8 (d)).

$$\dot{Q}_i = \dot{Q}_{i,cond} + \dot{Q}_{i,char\ rad} - \dot{Q}_{i,surf\ rad} + \dot{Q}_{i,amb\ rad} \tag{7}$$

In Eqs. (6) and (7) for flame standoff distance $\delta = L_{ph}$,

$\dot{Q}_{o,cond} \sim 2\pi\lambda_g L_{ph} \frac{(T_{fo}-T_{so})}{ln((R+L_{ph})/R)}$ and $\dot{Q}_{i,cond} \sim 2\pi\lambda_g L_{ph} \frac{(T_{fi}-T_{so})}{ln(R/(R-L_{ph}))}$ , when $R \gg L_{ph}$, $\dot{Q}_{o,cond} \sim 2\pi\lambda_g R(T_{fo} - T_{so})$ and $\dot{Q}_{i,cond} \sim 2\pi\lambda_g R(T_{fi} - T_{so})$. The other terms in Eqs. (6) and (7) are related to radiation heat transport, some of which require determination of radiation exchange view factors.

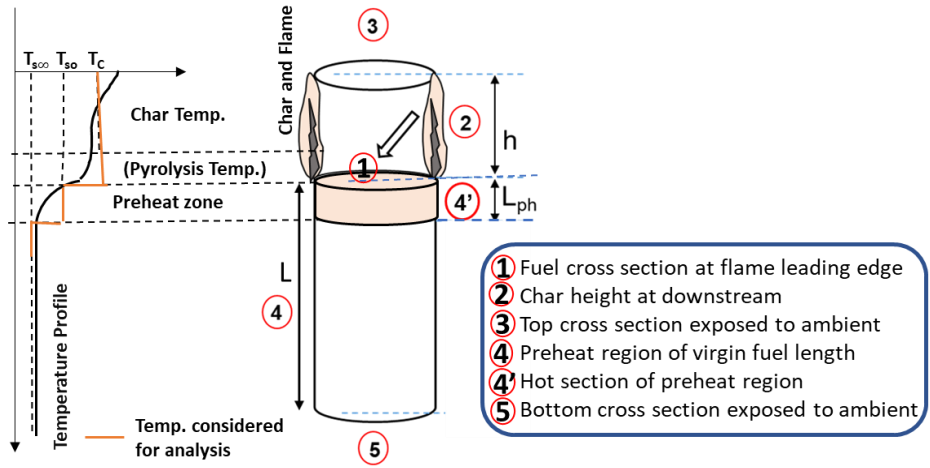

Fig. 9. Representation of different region of hollow cylindrical fuel for radiation heat feedback along with realistic and model temperature profiles.

Figure 9 illustrates various surfaces for estimating radiation heat exchange with the preheat region of the fuel. Figure 9 also shows the typical surface temperature profile and the assumed temperature profile along the length of the fuel. Surface 1 is the fuel cross section of radius 'R' at the flame leading edge, surface 2 is the char post pyrolysis, of height 'h' located downstream of flame leading edge is taken to be at temperature $T_C$, surfaces 3 and 5 are cross sections of radius 'R' at the top and bottom of the fuel exposed to the ambient, surface 4 is the preheat region or the unburnt virgin fuel of length 'L' and surface 4' is the hot section of the preheat region (at $T_{so}$) from which radiation heat loss is assumed to take place. It is to be noted here that surface 4' is a part of surface 4, is distinguished from surface 4. Surface 4 is used for incoming radiation from char whereas surface 4' is used for predicting radiation loss from the surface. This is done so because fuel surface temperature increases steeply close the flame leading edge due to conduction (see schematic of surface temperature profile in Fig 9). The radiative heating on the other hand in comparison to conduction is much smaller in magnitude and extends far upstream and decreasing gradually at progressively further upstream positions. For the outer surface radiation interaction in Eq. (6) is with the surrounding and can be written as

$$\dot{Q}_{0,surf\,rad} - \dot{Q}_{0,amb\,rad} = 2\pi R L_{ph} \epsilon_v \sigma T_{so}^4 - 2\pi R L_{ph} \epsilon_v \sigma T_\infty^4 \tag{8}$$

Here $\epsilon_v$ is the emissivity of virgin fuel surface and $\sigma$ is Stefan Boltzman constant.

In eq. (7) the terms $\dot{Q}_{i,char\,rad}$, $\dot{Q}_{i,surf\,rad}$, and $\dot{Q}_{i,amb\,rad}$ can be determined as

$$\dot{Q}_{i,char\,rad} = A_2 \epsilon_c \alpha_v \sigma T_C^4 F_{2-4} \sim A_2 \epsilon_c \alpha_v \sigma T_C^4 F_{2-1} F_{1-4} = A_1 \epsilon_c \alpha_v \sigma T_C^4 F_{1-2} F_{1-4} \tag{9}$$

Note that to obtain the view factor between surface 2 and 4, $F_{2-4}$ is quite involved and therefore approximated as $(A_1/A_2)F_{1-2}F_{1-4}$ considering two tandem cylinders with common interface of surface 1. Here, radiation energy from char (surface 2) incident on surface 1 is transmitted to cylinder below comprising of virgin fuel (surface 4) and surface 5. The introduction of intermediate surface 1 in the above approximation conserves total radiation energy transport but results in underestimation of radiation energy received by surface 4 and over estimation of radiation energy lost to environment. This in fact represents more closely the actual situation where char region is not a uniform surface and has large cracks and open spaces.

$$\dot{Q}_{i,surf\,rad} - \dot{Q}_{i,amb\,rad} = 2\pi R L_{ph} \sigma (T_{s0}^4 - T_\infty^4) \epsilon_v (F_{4'-1} + F_{4'-5})$$

$$\sim A_1 \sigma (T_{s0}^4 - T_\infty^4) \epsilon_v (F_{1-4'} + F_{1-4'}(1 - F_{1-4})) \tag{10}$$

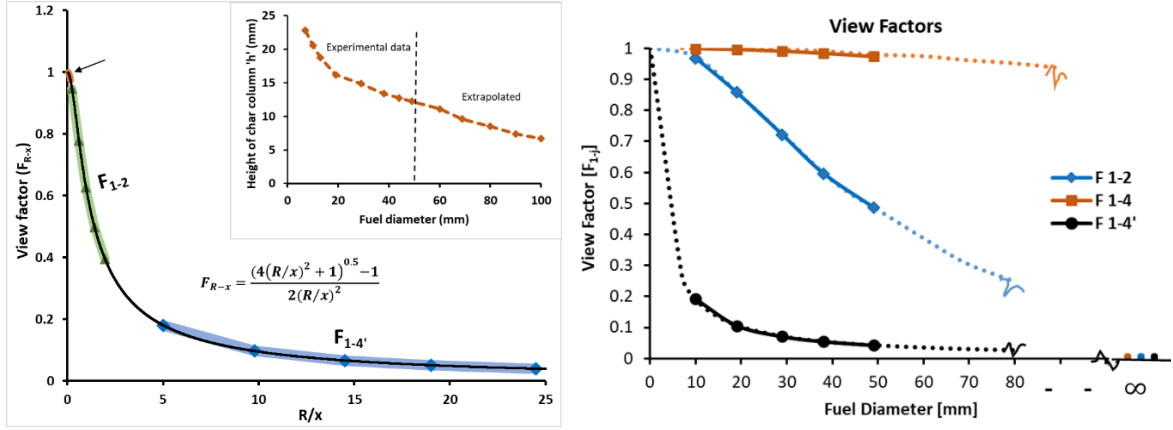

Fig. 10. (a) variation of view factors with respect to ('R/X') of fuel and char height with fuel diameter (Insert) (b) Variation view factors with respect to fuel diameter.

In the above equations, $A_j$ is the area of surface 'j' with surface emissivity, $\epsilon_j$ and absorptivity, $\alpha_j$. Here it assumed that $\epsilon_j = \alpha_j$. The radiation view factors after applying reciprocity and summation laws are finally represented in terms of radiation view factor, $F_{1-j}$ between surface 1 and one of the curved surfaces (j = 2, 4 or 4'). All view factors $F_{1-j}$ have the generic form $F_{R-x} = \frac{(4(R/x)^2+1)^{0.5}-1}{2(R/x)^2}$ [35]. Here $F_{R-x}$ represents view factor between top or bottom cross-section of a right circular cylinder of radius 'R' to the neighbouring curved surface of the cylinder of height '$x$'. Note that the view factor $F_{R-x}$ varies from 1 to 0 as '$R/x$' varies from '0' to '∞'. The variation of view factors $F_{1-2}$, $F_{1-4}$, and $F_{1-4'}$ with $R/x$ ranging from 0 to ∞ is shown in Fig. 10 (a). Also indicated in Fig. 10 (a) are thick solid line segments which represent the variation of these view factors over the present matrix of experiments. Note that here for representation $F_{1-4}$ is plotted for unburnt fuel length, L=150 mm and $F_{1-4'}$ is plotted for $L_{ph}$ of about 1 mm for flow speed of 20 cm/s. Since the char height, 'h' decreases significantly with fuel diameter, from 20.5 mm for fuel diameter of 10 mm to 12.5 mm for fuel diameter of 49 mm, the view factors $F_{1-2}$ are obtained for corresponding 'R/h' values. The variation of char height with fuel diameter for flow speed of 20 cm/s is also shown in the inset of Fig. 10(a). Note that the $L_{ph}$ and 'h' also vary with flow speed. Figure 10(b) presents variation of the three view factors with fuel diameter for the present set of experiments. The view factor between the char column and virgin fuel is $F_{1-2} \times F_{1-4}$ which varies from about 0.95 at 10 mm fuel diameter to about 0.38 at fuel diameter of 49 mm. The variation is primarily due to the view factor $F_{1-2}$ which varies from 0.9 to 0.4 over the same range of fuel diameter. $F_{1-4}$ on other hand remains close to 1. In the limit of R→∞ the radiation heating rate is same for the inner and outer surfaces, i.e. $\dot{Q}_{i,char\,rad} - \dot{Q}_{i,surf\,rad} + \dot{Q}_{i,amb\,rad} = \dot{Q}_{o,surf\,rad} - \dot{Q}_{o,amb\,rad}$ as it is for the planar fuels.

Using eq. (8) – (11) the expression for $\dot{Q}_o$ and $\dot{Q}_i$ in eq. (6) and (7) respectively can be written as

$$\dot{Q}_o = \lambda_g(T_{f0} - T_{so})2\pi R - 2\pi R \epsilon_s \sigma(T_{so}^4 - T_{s\infty}^4)L_{ph} \tag{12}$$

$$\dot{Q}_i = \lambda_g(T_{fi} - T_{so})2\pi R + A_1\epsilon_{ch}\alpha_v\sigma T_{ch}^4 F_{1-2}F_{1-4} - A_1\sigma(T_{s0}^4 - T_\infty^4)\epsilon_v\big(F_{1-4'} + F_{1-4'}(1 - F_{1-4})\big) \tag{13}$$

The flame spread rate can thus be obtained using eq. (5). The flame spread rate estimation procedure is discussed in the later section

**The overall equivalence ratio and the flame temperature.**

The hollow cylindrical fuel is placed inside a flow duct of cross section of size 12cm x 12cm. The incoming flow is divided into two parts, one going through the inner core of the hollow cylinder and other part is external to the hollow cylinder through space between the duct walls and the hollow cylinder. As the flame spreads the pyrolysate comes out from both inner and outer surfaces of the hollow cylindrical fuel. It is evident that the availability of oxygen for reaction in the inner core and in the region external to fuel surface are different. Depending on the diameter of the hollow cylindrical fuel and the imposed flow speed the overall equivalence ratio, '$\varphi_i$' in the core of the hollow cylindrical fuel may be on the rich side (or under ventilated). In such a situation the flame temperature ($T_{fi}$) in the core region will be lower than the flame temperature ($T_{fo}$) of the outer flame winglet which is close to stoichiometry (losses neglected). The flame temperature in the confined core region of a hollow cylindrical fuel of given diameter and flow condition can be estimated.

The mass rate of pyrolysate generated upon pyrolysis of cylindrical fuel of radius 'R' from a flame spreading at the rate of $V_f$ is $\dot{m}_f = 2\pi R \tau V_f \rho_s$. Here $\rho_s$ is the solid fuel density and $\tau$ is fuel thickness. It is assumed that half of the pyrolysate enters the inner core and remaining half leaves from the outer surface of the fuel. Now, the air mass flow rate in the inner core can be approximated as $\dot{m}_a = \pi R^2 U_\infty \rho_g$. Here $\rho_g$ is the gas density and $U_\infty$ is average gas flow speed. The overall equivalence ratio in the inner core region of the fuel can be written as

$$\varphi_i = \left(\frac{\dot{m}_f}{\dot{m}_a}\right) \bigg/ \left(\frac{F}{A}\right)_{stoich.} = \left(\frac{2\tau V_f \rho_s}{R U_\infty \rho_g}\right) \bigg/ \left(\frac{F}{A}\right)_{stoich.} \qquad (14)$$

Here $\left(\frac{F}{A}\right)_{stoich.}$ is the ratio of mass of fuel to mass of air at stoichiometric condition. One can note form eq. 14 that the equivalence ratio at inner core of the thin hollow fuel depends on fuel area density ($\tau \rho_s$), flow speed ($U_\infty$), fuel-air stoichiometric ratio $\left(\frac{F}{A}\right)_{stoich}$, flame spread rate ($V_f$), the fuel radius (R) and gas density ($\rho_g$). For the present study the fuel is cellulose ($C_6H_{10}O_5$) with area density of 18 gsm and $\left(\frac{F}{A}\right)_{stoich} = 0.196$.

The conditions in the inner core of the hollow cylindrical fuel may be fuel rich ($\varphi_i > 1$). In such case the flame temperature depends on the overall equivalence ratio. Therefore, the flame temperature in the core region of the fuel is expected to vary with hollow fuel diameter and the flow speed. Based on the equivalence ratio the effective temperature of the flame over inner region can be determined as

$$T_{fi} = T_\infty + \frac{\Delta H_R}{C_P(\varphi+\nu_S)} \quad \text{for } \varphi \geq 1 \quad (15)$$

For outer flame extension the overall equivalence ratio, '$\varphi_o$' is always < 1 as the maximum fuel diameter experimented here is 49 mm which is much smaller compared to the duct cross-section of 120 mm x 120 mm. Therefore, $T_{fo}$ and $T_{fi}$ for $\varphi \leq 1$ is taken to be same as for $\varphi = 1$

$$T_f = T_\infty + \frac{(\Delta H_R)}{C_P(1+\nu_S)} \quad \text{for } \varphi \leq 1 \quad (16)$$

In the above eq. (15) and (16) $T_{fi}$ and $T_{fo}$ are the flame temperatures of inner and outer flame extents, $T_\infty$ is the ambient temperature, $\Delta H_R$ is the heat of combustion for cellulose, $C_P$ is specific heat of the gas, $\varphi$ is equivalence ratio and $\nu_S = \left(\frac{\dot{m}_a}{\dot{m}_f}\right)$ is the overall air to fuel ratio.

**Flame spread rate estimation**

In the previous section eq. 5 gives an expression for estimating the flame spread rate. This requires an estimate of heating rate due to heat fluxes on the preheat region of the fuel. The heating rates (or integrated heat flux values) over the inner and outer surfaces of the hollow cylindrical fuel are given by eq. 6 and eq. 7. The radiation components of these heating rates are obtained from eq. 8-10 to yield final equations eq. 12 and eq. 13 for heating rates on the inner and outer surfaces of the hollow cylindrical fuel. The overall equivalence ratio inside the core of the hollow fuel is given by eq. 14 and the flame temperatures in the inner region and outer regions of the fuel are given by eq. 15 and eq. 16 respectively. The set of equations 5-16 are solved iteratively starting with a guess value of flame spread rate and fuel properties to obtain flame spread rate, view factors, radiation and conduction heat flux components, overall equivalence ratio in the inner region of the fuel, inner and outer flame temperatures. These calculations are performed for fuel diameters ranging from 7 mm to 100 mm and flow speed ranging from 10-30 cm/s.

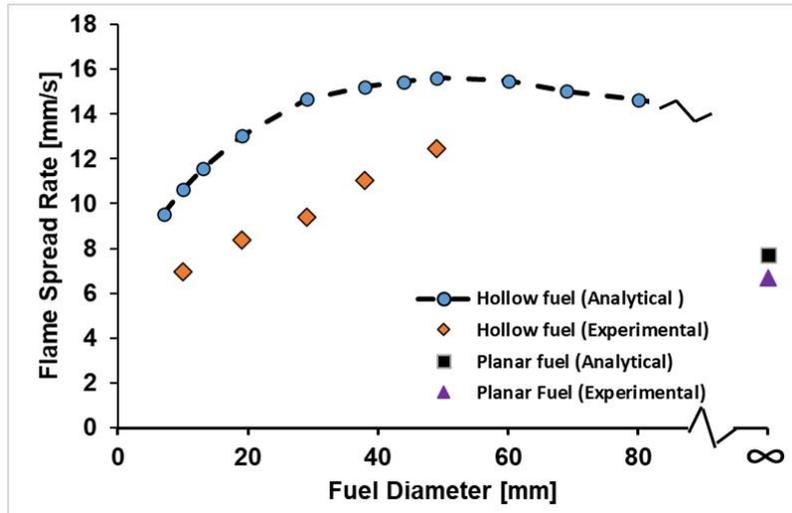

Fig. 11. Comparison of predicted flame spread rate over hollow cylindrical fuels from eq. 5 with experimental data for different fuel diameter in microgravity at 20 cm/s flow velocity. Experiment and predicted Flame spread rate for planar fuel at 20 cm/s flow is also shown for comparison.

Figure. 11. shows experimentally measured flame spread rates (diamond symbols) for different diameter of fuel and at an opposed flow speed of 20 cm/s. Also, shown in the figure is the predicted flame spread rate obtained from eq. (5) along with eq. (6)-eq. (12) using procedure described above. Flame spread rate for planar fuel is also marked in the plot for comparison. It is interesting to note that the experimental data shows a monotonic increase of flame spread rate with respect to fuel diameter for the range of fuel diameters (10 mm to 49 mm) studied in this work. Note that these flames spread rates are equal (at small fuel diameters) or higher than those for planar fuel (which is about 7 mm/s) under identical conditions. A similar trend was noted in the work of Itoh et al. [27] on downward flame spread over hollow cylindrical fuels in normal gravity. However, in the present study the flame spread rate predicted from the theoretical model for the same conditions shows a non-monotonic trend. The flame spread increases with fuel diameter, reaches a maximum at certain fuel diameter (here for 49 mm) and decreases with further increase in fuel diameter. In order to understand the non-monotonic trend in the flame spread rate with fuel diameter we look at eq. 5.

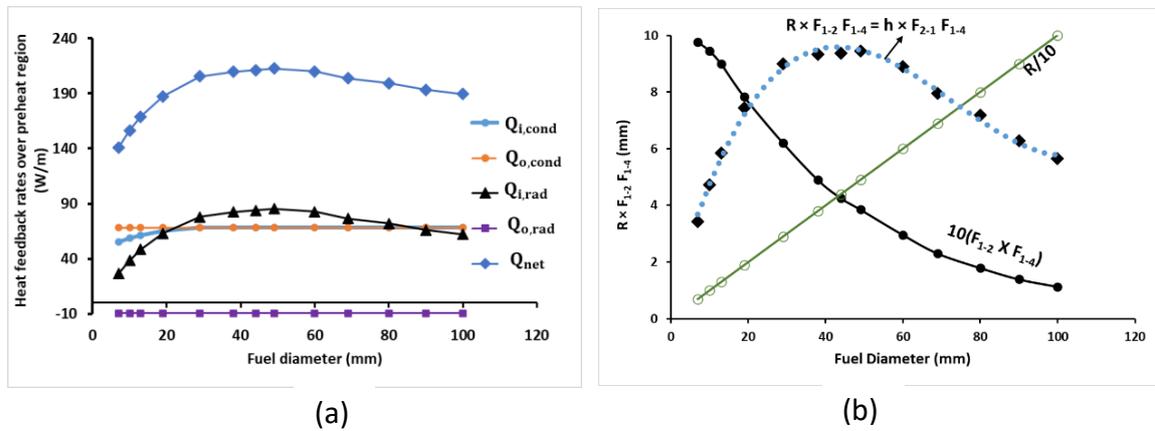

(a)            (b)

Fig. 12. (a) Variation of heat feedback components at inner surface and outer surface of fuel with fuel diameter at 20 cm/s opposed flow speed (b) Opposite trends of fuel preheat area and radiation view factor (char to preheat) contribution to non-monotonic trend of inner radiation feedback.

Since the flame spread rate in eq. (5) is directly proportional to net heat feedback to the preheat region on both inner and outer surfaces of the hollow cylindrical fuel, we next look at the heat feedback contribution to the preheat region from various sources and modes of heat transfer. The conduction and radiation heat feedback to the fuel preheat surface at inner and outer region of the hollow cylindrical, described in eq. (12) and eq. (13) are plotted against fuel diameter in Fig. 12 (a). The net heat feedback ($\dot{Q}_{net}$) coming to preheat region of hollow cylindrical fuel is the summation of conduction and radiation heat feedbacks at inner and outer surfaces of the fuel. As expected, following the trend of flame spread rate, the net heat feedback to the preheat region shows a non-monotonic trend with fuel diameter. One can note in Fig. 12 (a) that conduction ($\dot{Q}_{i,cond}$ and $\dot{Q}_{o,cond}$) is the dominant mode of heat transport to the fuel in the preheat region. Total conduction contribution (combined for inner and outer surfaces) is as high as 90 % of net heat feedback at small fuel diameter of 7 mm and reduces to about 60 % at maximum flame spread rate for fuel diameter of 49 mm. The contribution of conduction on inner ($\dot{Q}_{i,cond}$) and outer surfaces ($\dot{Q}_{o,cond}$) is nearly equal at large diameters, however, for small diameters contribution of conduction at inner surface decreases. Here this is noted for fuel diameters below 20 mm for flow speed of 20 cm/s. This decrease as we will see later is due to increase in overall equivalence ratio ($\varphi$) with reduces the flame temperature in the core region of the hollow fuel. While net radiation at the outer fuel surface is a mechanism by which the fuel loses heat from the fuel preheat region, radiation feedback to fuel surface in the inner core region ($\dot{Q}_{i,rad}$) of the fuel is the next prominent contributor (see Fig. 12(a)) to the total heat feedback to the preheat region. The radiation contribution is greatly reduced at small fuel diameter (here ~10 % for 7 mm fuel diameter) but contribution increases to about 40 % at the maximum flame spread rate and decreases gradually for further increase in hollow fuel diameter. It is interesting to note that the non-monotonic trend of inner radiation heat feedback ($\dot{Q}_{i,rad}$) with fuel diameter results in the non-monotonic trend of the total heat feedback ($\dot{Q}_{net}$) and thus the flame spread rate with fuel diameter over hollow cylindrical fuel. The non-monotonic trend of radiation feedback is explained next. There are two competing trends, one is the increase in fuel area which increases proportional to fuel diameter and second is the decreasing trend of view factor between the hot char region and the preheat region. These two trends with fuel radius are shown in Fig. 12(b) along with non-monotonic trend resulting due to the product of the opposing trends. It should be pointed out here that the high overall equivalence ratio at small fuel diameters also contributes to the flame spread rate, but the contribution here for opposed flow speed of 20 cm/s is small and may be considered secondary. In an earlier numerical study of flame spread over multiple planar fuel sheets (an array or large number of sheets) [23] a similar non-monotonic flame spread rate trend was observed with variation in fuel spacing. At small inter fuel spacing the flame spread increased with increase in fuel sheet spacing. This was primarily due to high overall equivalence ratio at small spacing which leads to lower flame temperatures on either side of fuel sheet. In the experiments of Avinash et al.[22] on downward flame spread over two parallel fuel sheets of width 2 cm, the flame spread rate increased from flame spread rates below that of a single sheet at small spacing of 0.5 cm to flame spread rate marginally higher than that of single sheet at inter fuel spacing of 3 cm. However, for 3 parallel fuel sheets a peak fame spread rate higher than single fuel sheet spread rate was noted as inter fuel spacing of 1.5 cm. In a similar study [36] a non-monotonic flame spread was reported for two parallel fuel sheets. Radiation from ember (char) was attributed for the observed behaviour. The effect of fuel width was seen to increase the peak flame spread rate. However, in their analysis effect of overall equivalence ratio was neglected and char height was assumed to be constant. It should be noted here that in case of parallel fuel sheets if there is entrainment of air from the sides, for small fuel widths this may alter the observed trend [22].

The above discussion so far has been for flow speed of 20 cm/s. The overall equivalence is expected to be influenced strongly by the imposed flow speed. Figure 13 shows variation of overall equivalence ratio in the inner core region of the fuel with fuel diameter for various flow speeds (Fig. 13(a)) and corresponding flame temperatures in the inner core region ($T_{fi}$) and outer regions ($T_{fo}$) (Fig. 13(b)). It is noted from Fig. 13 (a) that for any flow speed there is a maximum fuel diameter below which the overall equivalence ratio inside the fuel is greater than 1 (fuel rich). This maximum fuel diameter is higher for low flow speeds. For example, for the flow speed of 10 cm/s the overall equivalence ratio is rich up to the fuel diameter of 60 mm and this reduces to 20 mm for flow speed of 25 cm/s. Consequently, the flame temperature in the inner core region is lower than stoichiometric flame temperature in these fuel rich regions. On the outer surface where over all equivalence ratio is less than unity, the flame temperature is about the adiabatic value. Note, here in this work the heat loss from flame (both inner and outer) due to convection has been neglected, accounting for this loss can lower the predicted flame temperature below the adiabatic value depending upon the flow speed. Since the flame temperature at the inner core region is low due to fuel rich conditions for low flow speeds (for example here 10 cm/s), the flame spread rates are expected to be more severely affected than for higher flow speed of 20 cm/s noted in Fig. 13.

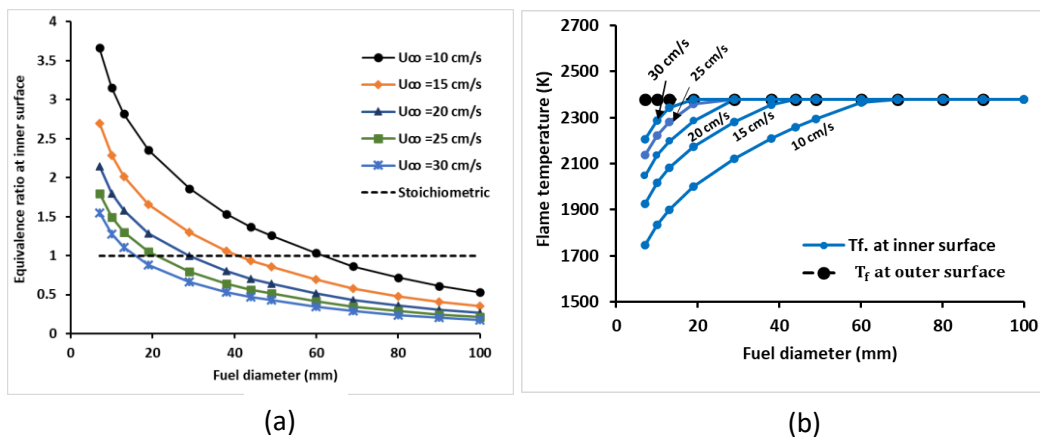

(a)          (b)

Fig. 13. Variation of (a) equivalence ratio at the inner core region of hollow cylindrical fuel and (b) flame temperature at inner and outer surfaces with fuel diameter at different opposed flow speeds

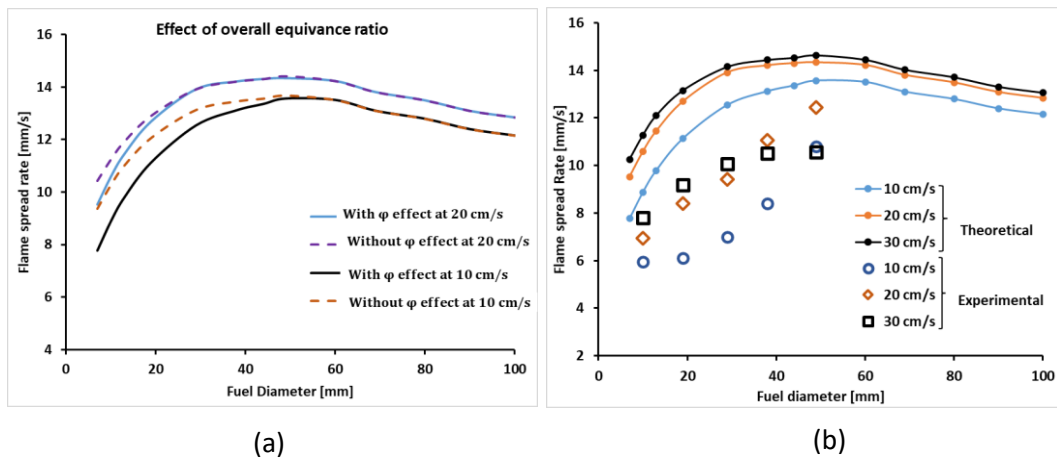

(a) (b)

Fig. 14 (a) Effect of overall equivalence ratio on predicted flame spread rate over hollow cylindrical fuel. (b) Comparison of predicted flame spread rate with experimental values at different opposed flow speed.

Figure 14 (a) compares predicted flame spread rate for two flow speeds of 10 cm/s and 20 cm/s for two conditions, one considering the effect of overall equivalence ratio in the inner region of the hollow fuel (solid lines) and other assuming adiabatic flame temperature in the inner region same as for the flame at outer surface. One can note the effect of overall equivalence ratio in the inner region is significant at low flow speed of 10 cm/s and effect extends up to fuel diameter of 60 mm. It should be pointed out here that in the present model flow inside the hollow cylinder is considered to be same as the imposed flow speed. It is expected that actual average flow speed in the inner core region would be only a fraction of the imposed flow speed. This will result in over equivalence ratio higher than predicted here and hence still lower flame spread rates. Figure 14 (b) shows variation of predicted and experimental flame spread rates with hollow cylindrical fuel diameter for various flow speeds. Both theory and experiments show that at small fuel diameter of 10 mm the flame spread rate increases with increase in flow speed. At this fuel diameter, contribution of radiation to flame spread rate is very small (Fig.12) and the increase in flame spread rate is directly attributed to decrease in overall equivalence ratio (less rich) at the inner region of the fuel. The experimental flame spread rate increases with fuel diameter for all the flow speeds with in the experimental matrix in this study. The predicted flame spread rate shows the same trend. However, for larger fuel diameters (here > 55 mm for all flow speeds) the flame spread rate decreases with further increase in fuel diameter. While this aspect is not captured in the present experiments, the flame spread rate for 30 cm/s opposed flow shows flattening of the flame spread rate curve at 38-49 mm fuel diameters. It is interesting to note that maximum flame spread rate is predicted at about nearly same fuel diameter irrespective of flow speed. This suggests that radiation feedback controls the flame spread rate and overall equivalence ratio is close to unity or less (Fig. 12 (a)). The crossover of experimental flame spread rate curves for various flow speeds was explained in the previous section to be due to non-monotonic trend of flame spread rates with flow speed.

The present analysis does not predict this trend as heat loss from the flame and effect of combustion kinetics are not modelled here.

**Conclusions**

In this work opposed flow flame spread process over thin cellulosic fuels in two geometrical configurations, namely hollow cylinder and planar is studied experimentally and theoretically. Experiments are carried out in microgravity environment in 1 atm. pressure and at 21% $O_2$ for a forced opposed flow speeds ranging from 10 - 30 cm/s. For cylindrical fuel, diameter was varied from 10 mm to 49 mm and for planar fuel width was varied from 10 mm to 40 mm. Images of the spreading flames were acquired to capture flames shapes and to obtain flame spread rates. A simplified analytical model for flame spread over thin hollow cylindrical fuel is developed to estimate the flame spread rate and explain the observed trends. The key findings of the study are summarized as follows.

1. The spreading flames over hollow cylindrical fuels in microgravity are less luminous but generally larger in size (both length and thickness) compared to their counterpart in normal gravity, except for small fuel diameters (here 10 mm) where flame size reduced in microgravity compared to normal gravity. Flames over planar fuels in microgravity were seen to be larger than normal-gravity flames. The luminosity or intensity of the flame in microgravity was seen to increase with increase in opposed flow speed.
2. The flame shape for hollow cylindrical fuel in microgravity changes from one only on the external surface at small fuel diameters to closed tip candle like flame at intermediate fuel diameters to and annular flame at large fuel diameters. The flames in normal gravity transitions to annular flame at relatively small fuel diameter between 10 mm and 19 mm.
3. For hollow cylindrical fuels, the flame length varies significantly with fuel diameter. In microgravity flame length increased with increase in fuel diameter. This trend is opposite of flame length decreasing with increase in fuel diameter was noted in normal gravity. For planar fuels there is no significant change in flame length with fuel width in both normal gravity and microgravity environments. This trend was noted for all the flow speeds.

4. Under-ventilation at small fuel diameters in the inner region of the hollow cylindrical fuel results in significant char formation in both microgravity and normal gravity environments. The char height decreases with increase in the diameter of hollow cylindrical fuel and with increase in flow speed. Planar fuels have char length always lower than flame height. Char length is longer in microgravity compared to normal gravity.

5. Conduction heat feedback from flame constitutes major fraction of total heating rate to the preheat region. Conduction heat feedback is affected by under-ventilation in the inner region of the hollow cylindrical fuels at small fuel diameters. Flame temperature in the inner region of fuel decreases as underventilation become severe (high equivalence ratio >1) which in turn reduces conduction feedback form the inner flame. Radiation heat feedback from the char to inner surface of the unburnt fuel is the other component of total heating rate to the preheat region. This radiation component has a non-monotonic increasing-decreasing variation with fuel diameter due to the exchange view factor. Estimated radiation contribution to total heating rate is maximum (here about 40%) at certain radius (here about 50 mm) and insignificant at small diameters (here less than 10% for fuel diameter of 10 mm).

6. The microgravity flame spread rate over thin hollow cylindrical fuels has a non-monotonic increasing-decreasing trend with fuel diameter. This trend is primarily a consequence of radiative heating of preheat region by hot char. The overall equivalence ratio or underventilation in the inner core region also contributes to increase in flame spread with fuel diameter. However, the effect of overall equivalence ratio is limited to small diameters and low flow speeds. Therefore, the fuel diameter at which maxima in flame spread rate is expected to be governed by this radiation exchange. The flame spread rate of hollow cylindrical fuel is either greater or equal (in case of small diameter at low flow speed) to flame spread over planar fuel. However, the flame spread rate over hollow cylindrical fuel will be lower than planar fuel spread rate at still smaller fuel diameter. Planar fuel flame spread rate showed no significant change in flame spread rate.

7. The micro-gravity flame spread rate over hollow cylindrical fuel as well as planar fuel exhibit a non-monotonic increasing-decreasing trend with respect to opposed flow speed. The flow speed at which maxima in flame spread rate occurs depends of the fuel diameter. For small fuel diameter a maxima will occur at higher flow speeds compared to fuel with larger diameter. In the present study the smaller fuel diameter (here 10 mm and 19 mm) and small width planar fuel (10mm) show a monotonic increasing trend with increase in opposed flow speed, whereas large diameters (> 19 mm) exhibit non-monotonic trend with in the flow speed range of this study.

**Acknowledgement**

This research was sponsored by Space Applications Centre (SAC), Indian Space Research Organization (ISRO), with Ms. Payal Sharma and Mr. Akash Gupta as technical monitors. We are thankful to OGGs-STSI program of Hokkaido University which provided the first author an opportunity to visit Space Utilization laboratory and having a wonderful technical discussion.YK thanks the support from Japan Society for the Promotion of Science (JSPS KAKENHI grant number P23K13258). The authors are grateful to Mr. Niketh, Mr. Abhishek, Mr. Abhinandan and Mr. Naveen for their great help during the experiments in drop tower laboratory, IIT Madras.